\documentclass[twocolumn,english,pra,showpacs,preprintnumbers]{revtex4-1}

\usepackage[utf8]{inputenc}
\usepackage{color}
\usepackage{amsmath}
\usepackage{amssymb}
\usepackage{graphicx}
\usepackage{tikz}
\usepackage{url}

\DeclareUnicodeCharacter{0308}{@}
\DeclareUnicodeCharacter{2212}{-}

\makeatletter

\@ifundefined{textcolor}{}
{%
 \definecolor{BLACK}{gray}{0}
 \definecolor{WHITE}{gray}{1}
 \definecolor{RED}{rgb}{1,0,0}
 \definecolor{GREEN}{rgb}{0,1,0}
 \definecolor{BLUE}{rgb}{0,0,1}
 \definecolor{CYAN}{cmyk}{1,0,0,0}
 \definecolor{MAGENTA}{cmyk}{0,1,0,0}
 \definecolor{YELLOW}{cmyk}{0,0,1,0}
}

\usepackage{textcomp}
\usepackage{dcolumn}
\usepackage{bm}
\usepackage[right]{eurosym}
\usepackage{float}
\usepackage[english]{babel}
\usepackage{blindtext}

\makeatother

\usepackage{babel}
\begin{document}

\title{Inelastic scattering of photoelectrons from He nanodroplets}

\author{M. V. Shcherbinin$^{1}$, F. Vad Westergaard$^{1}$, M. Hanif$^{1}$, S. R. Krishnan$^{2}$, A. C. LaForge$^{3}$\email{Present address: Department of Physics, University of Connecticut, Storrs, Connecticut, 06269, USA}, R. Richter$^{4}$, T. Pfeifer$^{5}$, and M. Mudrich$^{1}$\email{mudrich@phys.au.dk}}

\affiliation{$^{1}$Department of Physics and Astronomy, Aarhus University, 8000 Aarhus C, Denmark}

\affiliation{$^{2}$Department of Physics, Indian Institute of Technology, Madras, Chennai 600 036, India}

\affiliation{$^{3}$Department of Physics, University of Connecticut, Storrs, Connecticut, 06269, USA}

\affiliation{$^{4}$Elettra-Sincrotrone Trieste, 34149 Basovizza, Trieste, Italy}

\affiliation{$^{5}$Max-Planck-Institut f{\"u}r Kernphysik, 69117 Heidelberg, Germany}

\begin{abstract}
We present a detailed study of inelastic energy-loss collisions of photoelectrons emitted from He nanodroplets by tunable extreme ultraviolet (XUV) radiation. Using coincidence imaging detection of electrons and ions, we probe the lowest He droplet excited states up to the electron impact ionization threshold. We find significant signal contributions from photoelectrons emitted from free He atoms accompanying the He nanodroplet beam. Furthermore, signal contributions from photoionization and electron impact excitation/ionization occurring in pairs of nearest-neighbor atoms in the He droplets are detected. This work highlights the importance of inelastic electron scattering in the interaction of nanoparticles with XUV radiation.
\end{abstract}


\date{\today}

\maketitle

\section{Introduction}
Energetic electrons created inside a condensed phase system primarily interact with the individual atoms of that substance and their scattering can be predicted quite accurately. In contrast, low-energy electrons interact with the whole molecular network, and their scattering is currently too complex to make accurate predictions. In particular, inelastic and elastic scattering of slow electrons in water is lacking a complete understanding ~\cite{Alizadeh:2015}.
Therefore, precise measurements using simple model systems can add to the fundamental understanding of electron scattering in the condensed phase. Here, we present experiments with helium (He) nanodroplets which feature (i) an extremely simple electronic structure of the He constituent atoms, (ii) a homogeneous, superfluid density distribution~\cite{Harms:1998,Toennies:2004}, and (iii) a high electrophobicity, which facilitates the emission of slow electrons out of the droplets and thus allows for their sensitive detection.

He nanodroplets are commonly regarded as ``ideal spectroscopic matrices'' providing a transparent, cold, and weakly perturbing environment for the spectroscopy of embedded molecular species~\cite{Toennies:2004,Stienkemeier:2006}. They can be seen as flying nano-cryostats in which individual molecules are isolated and cooled upon pickup of a single molecule per droplet. When doping two or more molecules per droplet at elevated vapor pressure of the dopant species, aggregation into ultracold complexes with sometimes unusual configurations occurs~\cite{Nauta2:1999,Nauta3:2000,Przystawik:2008}. 
Thus, applying advanced spectroscopic techniques to dopants inside He nanodroplets may open new ways of probing the structure and dynamics of unconventional molecular complexes and nanoparticles~\cite{MuellerPRL:2009,Goede:2013,Thaler:2015,Bruder:2018}.

While photoionization combined with ion detection has proven to be a useful technique to probe neutral and cationic molecules~\cite{Mudrich:2004,Przystawik:2006,Loginov:2008,Smolarek:2010,Lackner:2014,Mudrich:2014}, the powerful technique of photoelectron spectroscopy is less established. Most photoelectron studies have been carried out using resonant two-photon ionization by nanosecond laser pulses, where local rearrangement of the He solvation shell around the intermediate excited dopant may impact the spectra~\cite{Radcliffe:2004,Loginov:2005,Peterka:2006,Przystawik:2007,Vangerow:2014}. A few studies of photoelectron spectra using one-photon ionization by XUV radiation have been reported for pure~\cite{Peterka:2003,BuchtaJCP:2013,Ziemkiewicz:2015,Shcherbinin:2017} and doped He droplets~\cite{Wang:2008,Buchta:2013,LaForgePRL:2016,Shcherbinin:2018}. However, in the latter case, dopants were ionized indirectly either by Penning ionization or by charge transfer from ionized He nanodroplets. Therefore, no information about the dopants was extracted from the measured electron spectra.

One difficulty in performing ultraviolet photoelectron spectroscopy (UPS) or even x-ray photoelectron spectroscopy (XPS) of doped He nanodroplets is that the emitted photoelectrons interact with the He matrix on their way from the photoionized dopant to the detector. This can lead to unwanted loss of angular information (anisotropy), to distortions of the electron spectra, or even to electron-ion recombination. Note that He droplets are even capable of trapping electrons when bombarding the droplets with electrons of several eV of kinetic energy~\cite{Henne:1998}. Anionic He droplets as well as anionic atomic and molecular He ions are also detected at electron impact energies around 22 and 44~eV~\cite{Mauracher:2014}. At these energies, the impinging electron excites one or even two He atoms inside the droplet and subsequently attaches to one excited He$^*$ atom. Similarly, by photoionizing He droplets at photon energies $h\nu>44~$eV, we have previously found indications that the photoelectron undergoes inelastic collisions with the He thereby losing around 22~eV of kinetic energy~\cite{BuchtaJCP:2013}. In contrast, the spectra of electrons originating from ionization and correlated decay processes revealed only weak perturbations by the He droplets~\cite{Shcherbinin:2017}. The present study is devoted to inelastic scattering of photoelectrons with He nanodroplets. We resolve individual components of the electron energy-loss spectra measured at various photon energies. From the analysis of peak positions and amplitudes, we infer various inelastic scattering scenarios. 



\section{Methods}
The experiments are performed using a He nanodroplet apparatus combined with a velocity map imaging photoelectron-photoion coincidence (VMI-PEPICO) detector at the GasPhase beamline of Elettra-Sincrotrone Trieste, Italy. The apparatus has been described in detail elsewhere~\cite{Buchta:2013,BuchtaJCP:2013}. Briefly, a beam of He nanodroplets is produced by continuously expanding pressurized He (50~bar) of high purity out of a cold nozzle (10-28~K) with a diameter of 5~$\mu$m into vacuum. At these expansion conditions, the mean droplet sizes range between $\langle N\rangle = 700$ and $\sim 5\times 10^6$ He atoms per droplet. Further downstream, the beam passes a mechanical beam chopper used for discriminating droplet-beam correlated signals from the background. 

In the detector chamber, the He droplet beam crosses the synchrotron beam perpendicularly in the center of the VMI-PEPICO detector. By detecting either electrons or ions with the VMI detector in coincidence with the corresponding particles of opposite charge on the TOF detector, we obtain either ion mass-correlated electron VMIs or mass-selected ion VMIs. Kinetic energy distributions of electrons or ions are obtained from the VMIs by Abel inversion~\cite{Dick:2014}. The energy resolution of the electron spectra obtained in this way is $\Delta E/E\gtrsim 5$\%. In this study, the XUV photon energy is tuned in the range $h\nu=44$-$64$~eV. Ion mass distributions from the He droplet beam recorded at these photon energies contain a series of cluster masses He$^+_n$, $n=1,\,2,\,3\dots$; but by far the most abundant fragments are He$^+$ and He$_2^+$~\cite{BuchtaJCP:2013}. All electron and ion spectra discussed in this work are measured at a He nozzle temperature of $T=14$~K which corresponds to a mean number of atoms per droplet of $N=23,000$, unless otherwise specified.

\section{Results and discussion}
Photoelectron spectra and photoelectron angular distributions of He nanodroplets near the ionization threshold have been studied before~\cite{Peterka:2003,Peterka:2007,BuchtaJCP:2013}. In the experiment using PEPICO-VMI recorded in coincidence with He$^+$, both the electron energy and the anisotropy was found to match that of the free He atom ($\beta =2$)~\cite{BuchtaJCP:2013}. In contrast, electrons measured in coincidence with He$_2^+$ were slightly upshifted in energy and their anisotropy was reduced to $\beta\approx 1$. Therefore, we concluded that He$^+$ ions created near the ionization threshold predominantly originate from free He atoms accompanying the He droplet beam, whereas He$_2^+$ and larger molecular ions He$_n^+$, $n>2$ are ejected from ionized He nanodroplets. 

\begin{figure}
	\centering \includegraphics[width=0.48\textwidth]{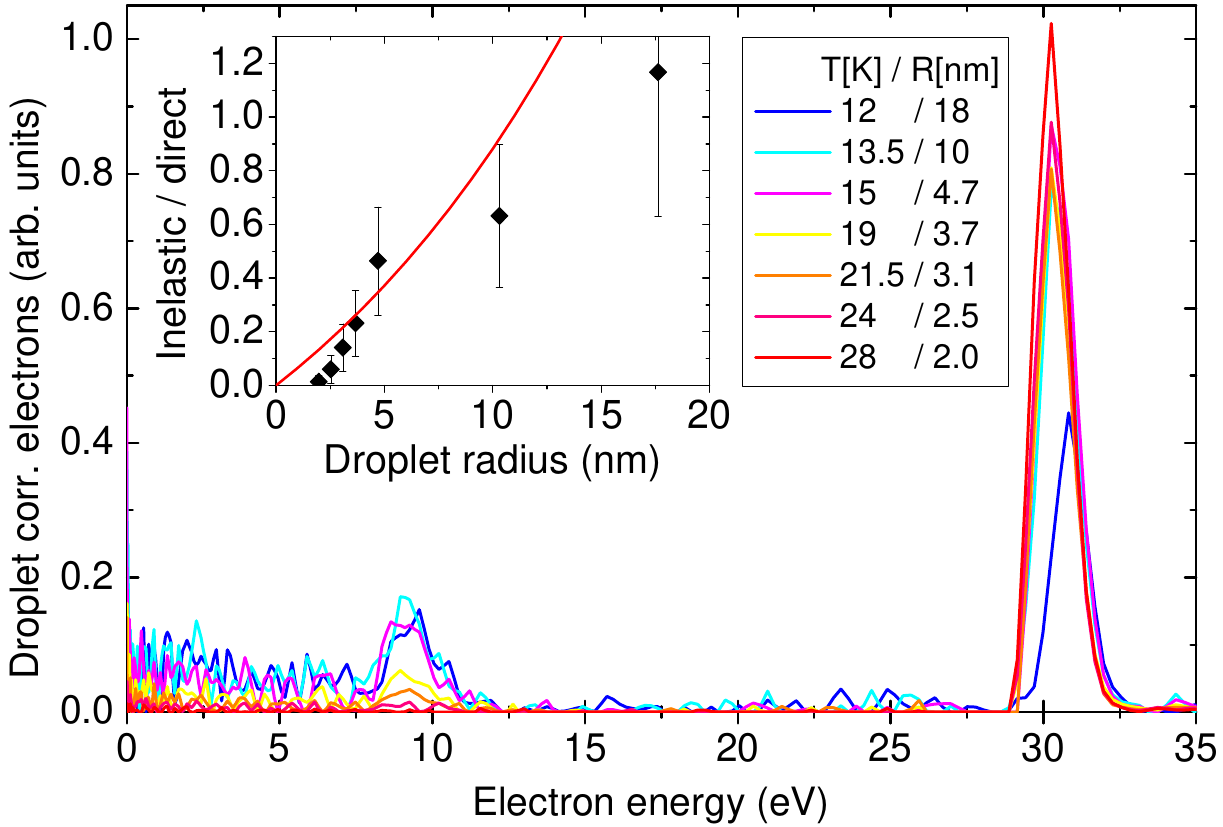}
	\protect\caption{\label{fig:temp} Spectra of total electrons emitted from He nanodroplets at a photon energy $h\nu = 55$~eV. The mean size of the droplets is varied by changing the temperature of the He nozzle. The inset shows the ratio of integrals over the low-energy peak (inelastic scattering) vs. the high-energy peak (direct photoemission).}
\end{figure}
When the photon energy is tuned to $h\nu\gtrsim 44$~eV, additional peaks and broad features appear in the electron spectra which are shifted to lower energies by $\gtrsim 18.5$~eV, see Fig.~\ref{fig:temp}. The peaks are due to inelastic scattering of the primary photoelectron with the He nanodroplets thereby exciting He atoms into excited states or into the ionic continuum. The fraction of inelastically scattered electrons with respect to those emitted directly with a kinetic energy around $h\nu-E_i$ rises from 0 up to $>1$ when the mean radius $R$ of the He droplets is increased from 2 to about 20~nm by lowering the temperature of the He nozzle from 28 to 12~K, see inset in Fig.~\ref{fig:temp}. Here, $E_i=24.59$~eV is the atomic ionization energy of He. The red line depicts the estimated fraction of inelastic collisions, $\exp(\sigma n_\mathrm{He}R)-1$. Here, $\sigma=0.29$~\AA$^2$~is the total inelastic collision cross section at $h\nu = 55$~eV~\cite{Ralchenko:2008}, and $n_\mathrm{He}=0.0218$~\AA$^{-3}$~is the density of He atoms in He nanodroplets~\cite{Harms:1998}. The reasonable agreement of our simple estimate with the experimental data confirms our interpretation. Accordingly, the mean free path of electrons with a kinetic energy of $h\nu-E_i=30.41$~eV in He nanodroplets due to inelastic scattering is $1/(\sigma n_\mathrm{He})=15.8$~nm. When the mean He droplet size is further increased to $>100$~nm (not shown), this ratio of inelastic collisions versus directly emitted electrons further rises to $>10$, but an additional feature eventually dominates the electron spectrum which will be discussed elsewhere.

\subsection{Electron energy-loss spectra}
\begin{figure}
	\centering \includegraphics[width=0.45\textwidth]{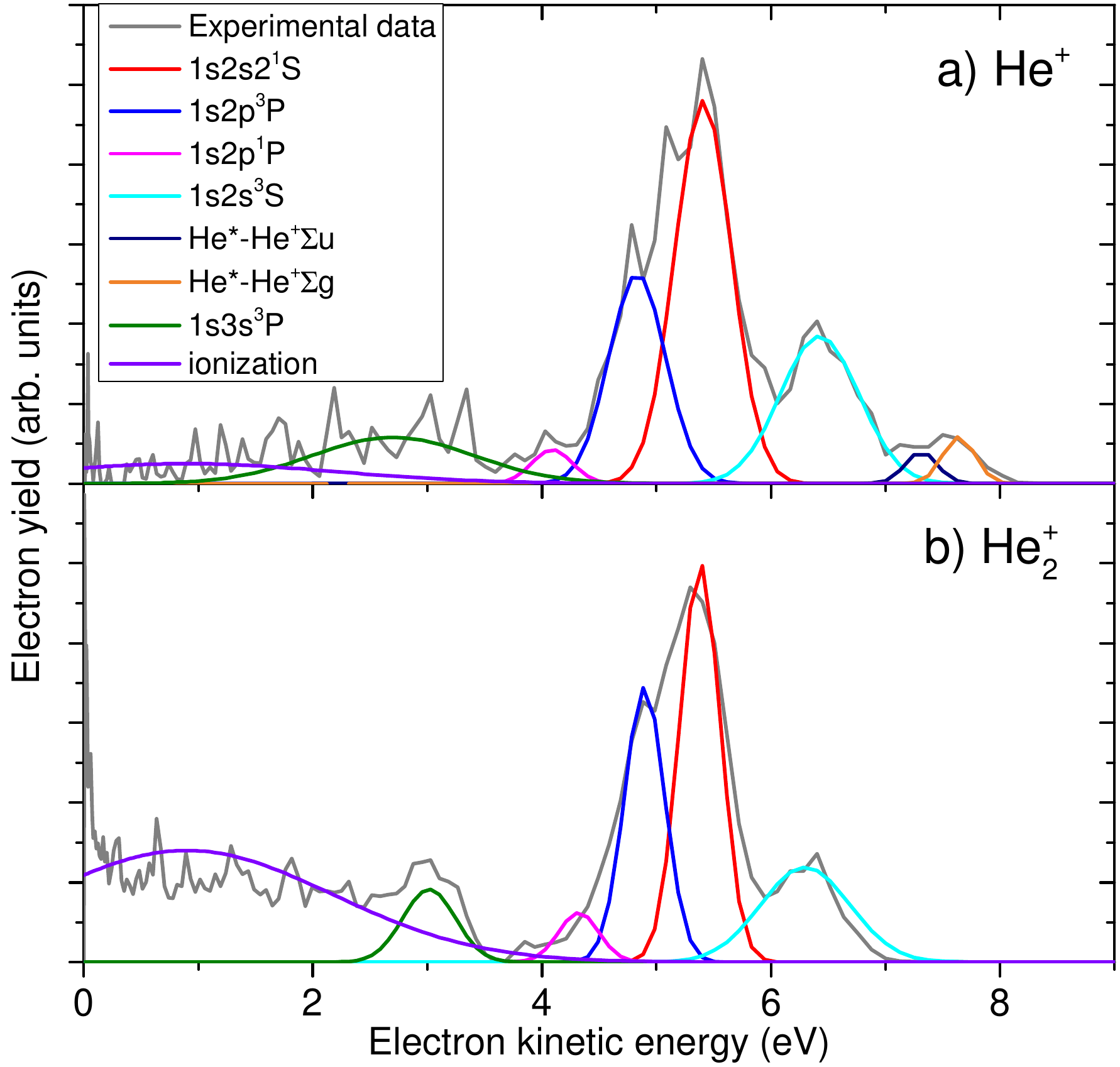}
	\protect\caption{\label{fig:EELS} Photoelectron spectra recorded in coincidence with He$^+$ and He$_2^+$ ions at a photon energy $h\nu = 51$~eV. Only the low-energy part is shown, where a multi-peak structure is generated by inelastic scattering of the primary photoelectron with He atoms thereby exciting them into various excited states. }
\end{figure}
\begin{figure}
	\centering \includegraphics[width=0.5\textwidth]{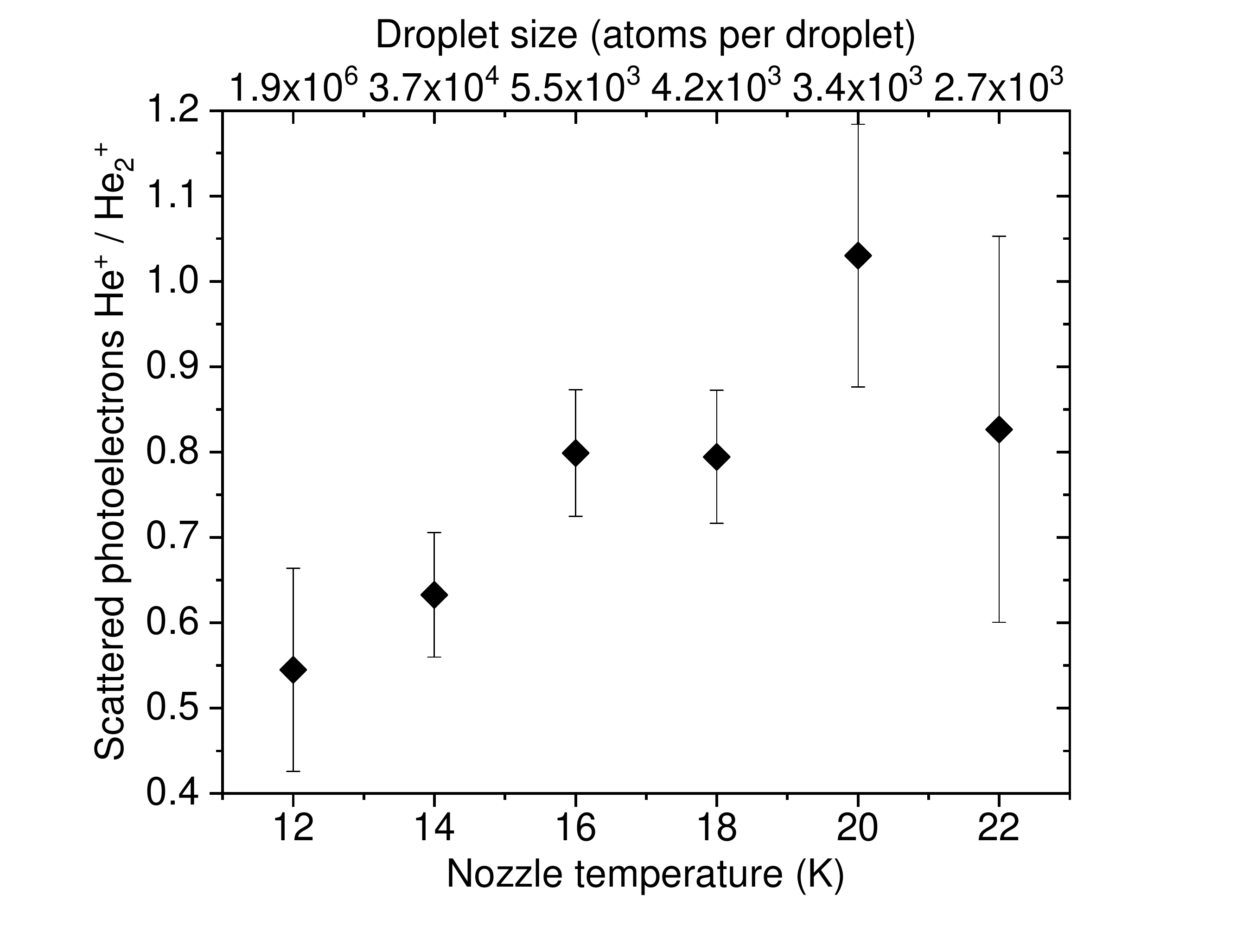}
	\protect\caption{\label{fig:t-dependance} Ratio of inelastically scattered electrons detected in coincidence with He$^+$ versus He$_2^+$ as a function of the He nozzle temperature (bottom axis) which controls the He nanodroplet size (top axis).}
\end{figure}

Typical raw spectra recorded at higher resolution and in coincidence with He$^+$ and He$_2^+$ are shown in Fig.~\ref{fig:EELS} a) and b), respectively. The photon energy is set to $h\nu = 51$~eV. In accordance with our previous interpretation, we attribute those electrons detected in coincidence with He$^+$ atomic ions to photoionization of mainly free He atoms. The fact that very similar electron energy loss spectra are measured in coincidence with He$^+$ and He$_2^+$ ions indicates that electrons are emitted from the free He atoms in the vicinity of the He droplets where they scatter inelastically. This interpretation is supported by the dependence of the yield of inelastically scattered electrons detected in coincidence with He$^+$ in proportion to those detected in coincidence with He$_2^+$, depicted in Fig.~\ref{fig:t-dependance}. As the temperature of the He nozzle is increased from 12 to 22~K and thus the mean droplet size decreases from about $1.9\times 10^6$ to 2,700 He atoms per He droplet, the ratio of inelastically scattered electrons (integral over electron spectra from 0 to 8~eV) in coincidence with He$^+$ versus He$_2^+$ rises by nearly a factor of 2.

The smooth bell shaped curves in Fig.~\ref{fig:EELS} depict gaussian functions which are simultaneously fitted to the experimental spectrum. Due to the limited quality of the experimental data, only 8 peaks can be identified with high confidence. We attribute them to the lowest excited states 1s2s$^3$S up to 1s3p$^1$P as well as the ionization continuum. As these states have the highest impact excitation cross sections~\cite{Ralchenko:2008}, they are expected to dominate the spectrum. From the peak positions of the fit curves, we infer the energy loss which equals the excitation energy of the respective state (see legend). The peak integrals correspond to the relative probabilities of exciting the various states by electron impact, see Sec.~\ref{sec:xsec}. 

The broad feature in Fig.~\ref{fig:EELS} b) reaching from zero up to 3.5~eV electron energy is due to electrons created by electron impact ionization of He. The flat structure of this feature indicates that the energy is shared between the two electrons according to a uniform distribution function, in accordance with previous findings~\cite{Cvejanovic:1974}. Note that in our experiment, we detect only one of the two electrons due to the finite deadtime of the detector. While the peaks in Fig.~\ref{fig:EELS} a) and b), corresponding to excited states, are similar in positions and amplitudes, the distribution assigned to impact ionization is more pronounced in the coincidence measurement with He$_2^+$.

Another significant difference between the spectra in coincidence with He$^+$ and with He$_2^+$ is the occurrence of two small peaks at an electron energy between 7 and 8~eV in the He$^+$ electron spectrum [Fig.~\ref{fig:EELS} a)], not present in the He$_2^+$ electron spectrum. These peaks are present in all He$^+$-correlated spectra except for those recorded at expansion conditions when large He droplets ($N>10^5$) are formed (nozzle temperature $T<14$~K).

\begin{figure}
	\centering \includegraphics[width=0.45\textwidth]{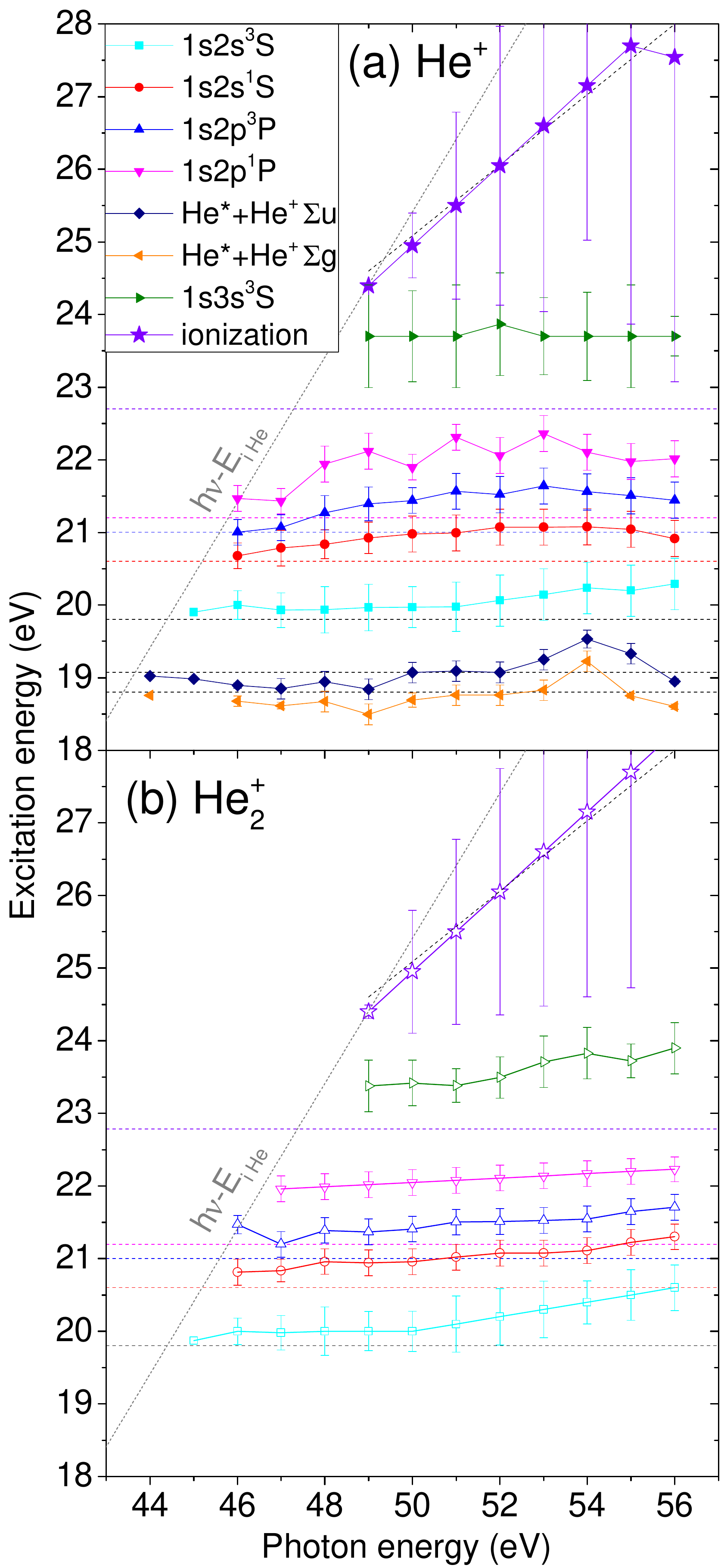}
	\protect\caption{\label{fig:PeakEnergies} Compilation of the fitted peak positions $E_p$ in electron spectra recorded in coincidence with He$^+$ a) and with He$_2^+$ b) as a function of the photon energy $h\nu$. The results are represented as energies of the He states excited by electron impact, $E^*=h\nu - E_i - E_p$, where $E_i$ denotes the atomic ionization energy of He.}
\end{figure}
To get an overview of the results at all measured photon energies ranging from 44 up to 56~eV, we plot in Fig.~\ref{fig:PeakEnergies} the energies of the He states excited by electron impact, $E^*=h\nu - E_i - E_p$, obtained from the fitted peak positions $E_p$. The error bars indicate the widths of the fitted peaks (standard deviation). 
The choice of the atomic ionization energy as the value of $E_i$ is well justified when analyzing the He$^+$ coincidence electron spectra which are primarily due to photoionization of free He atoms. As for the He$_2^+$ coincidence data, the corresponding value of $E_i$ may be slightly reduced by about 0.1~eV as observed in photoelectron spectra recorded near-threshold~\cite{Peterka:2007,BuchtaJCP:2013}. However, for the sake of consistency with the representation of the He$^+$ data, and since the shift of $E_i$ is small compared to the resolution of our spectrometer, we use the same value $E_i=24.59$ throughout.

The horizontal dashed lines in Fig.~\ref{fig:PeakEnergies} represent the He excited state energies $E^*$ for the He atomic values $E_p$~\cite{NIST}. The slanted dashed lines tangent to the onset of the data points at low photon energies represent the highest possible energy that the primary photoelectron can transfer by exciting a He atom, $h\nu-E_i$. The slanted dashed lines that nearly match the fitted positions of the broad impact ionization feature represents the linear function $h\nu-3E_i/2$, that is the expected energy of electrons created by impact ionization when assuming equal energy sharing between the two electrons. 

Overall we find a good correspondence between the fitted peak energies and the literature values~\cite{NIST}. However, the experimental atomic excitation energies are systematically up-shifted in energy by 0.2-1~eV, where the higher excited states are up-shifted more than the lowest He excited state 1s2s$^3$S. We attribute this up-shifting to the repulsive interaction of He$^*$ excited atoms (assuming prompt localization of the excitation on one atom) with the surrounding ground state He atoms inside the He nanodroplet. This concept is commonly adopted to explain the broad, blue-shifted features in photoabsorption spectra of He nanodroplets~\cite{Joppien:1993,Haeften:2011,Kornilov:2011}. For the higher excitations into 1s$n\ell$-states with principal quantum number $n=3,\,4$ and $\ell=0,\,1,\,2$, even the ejection of free Rydberg atoms was observed~\cite{Kornilov:2011}. In our experiment we find an average up-shift of the excitation energy of the 1s2p$^1$P-state of $\Delta E_\mathrm{^1P}=0.8\pm 0.2$~eV,
which matches the blue-shift of the absorption peak very well~\cite{Joppien:1993}. In this way, we can specify the energetic up-shift for the optically forbidden states 1s2s$^3$S and 1s2p$^3$P to $\Delta E_\mathrm{^3S}=0.35\pm 0.1$~eV and $\Delta E_\mathrm{^3P}=0.45\pm 0.2$~eV, respectively. Likewise, we obtain values for the full widths at half maximum (FWHM) for each peak. These values range from $0.7\pm 0.1$~eV (1s2s$^{1}$S and 1s2p$^{1,\,3}$P), $0.9\pm 0.2$~eV (1s2s$^{3}$S) to $1.6\pm 0.6$~eV (1s3p$^1$P). The peak widths for the optically allowed states are in good agreement with those measured by photoabsorption spectroscopy. 

\begin{figure}
	\centering \includegraphics[width=0.48\textwidth]{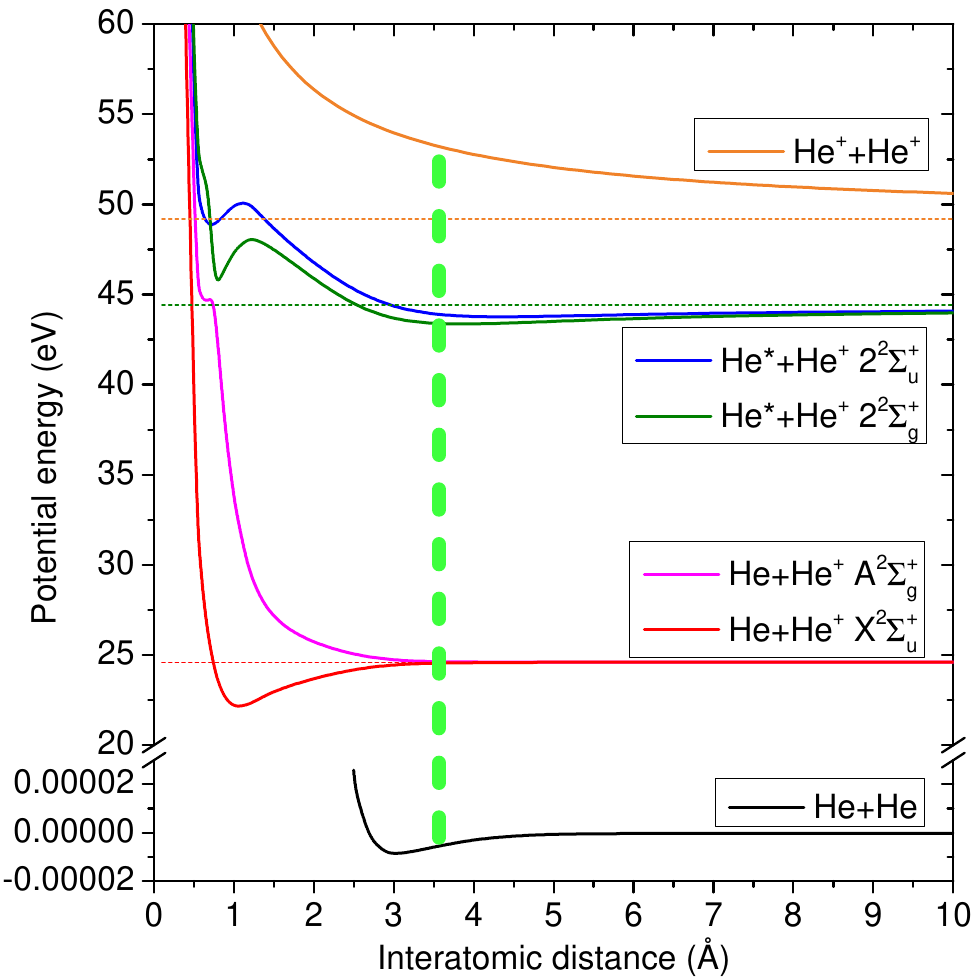}
	\protect\caption{\label{fig:Potentials} Selected potential energy curves for the He$_2$ ground state up to the doubly ionized state He$^+ + $He$^+$, taken from~\cite{Ackermann:1991,Janzen:1997}. The thick vertical dashed line indicates the average distance between He atoms inside He nanodroplets. }
\end{figure}
\subsection{Nearest-neighbor excitation and ionization by electron collision}
In addition to the peaks corresponding to atomic excitations of He, we find two small peaks at energies about 1~eV below the lowest excited atomic level 1s2s$^3$S in the He$^+$ coincidence spectra [Fig.~\ref{fig:EELS} a)]. How can an electron lose less energy than the lowest He excitation in an inelastic collision with a He nanodroplet? We have mentioned that the presence of neutral He atoms around the impact-excited He atom can only cause an up-shift of the energy of the low-lying levels. The interpretation we propose is based on the peculiar shape of potential curves for a pair of He atoms where one is excited and the other is ionized, see the green and blue lines in Fig.~\ref{fig:Potentials}. In the range of most probable He-He interatomic distances inside He nanodroplets, around 3.6~\AA~\cite{Peterka:2007}, the lowest two potential curves $2^2\Sigma_{g,u}^+$ correlating to the pair of atoms He$^* + $He$^+$ feature a shallow well with a depth of 0.6 and 1.0~eV with respect to the He$^*$(1s2s$^3$S)$ + $He$^+$ atomic asymptote, respectively. Thus, we assume that the primary photoelectron undergoes an inelastic collision with a neighboring He atom whose excitation energy is down-shifted by the presence of the nearby photoion. We mention that this process resembles the well-known shake-up and knock-up processes in an atom or a molecule, where an electron is emitted by the absorption of an energetic photon and simultaneously the remaining photoion is electronically excited~\cite{Yarzhemsky:2016}. Shake-up, which is driven by electron correlation, is less likely to play a role here given the large distance between two He atoms~\cite{Kirill:2018}. However, we cannot exclude its contribution to the measured signal given that more than one atom surrounds the ionization center which may lead to a collective enhancement. 

According to our interpretation of the two additional peaks in terms of He$^* + $He$^+$ molecular excitations, we have added dashed horizontal lines at excitation energies of 18.8 and 19.2~eV in Fig.~\ref{fig:PeakEnergies} a). The good agreement of these values with the experimental peak energies confirms our model. The missing up-shift for these states is due to the fact that now the nearest neighbor of the He$^*$ is the He$^+$ photoion which down-shifts the level. This is in contrast to the more frequent cases where the photoion is at some distance away from the He$^*$, and the nearest neighbor to the He$^*$ is a neutral He atom causing an up-shift of the He$^*$ level, as discussed above. Note that there may be more down-shifted features due to molecular excitations correlating to the higher-lying excited atomic levels superimposing on the electron energy-loss spectrum. However, given their low relative amplitudes and the limited quality of our data, we cannot unambiguously identify any.

But why are the two He$^*$-He$^+$ molecular features observed only in coincidence with He$^+$ atomic ions? Clearly, the combined process of photoionization and scattering on the next neighbor occurs inside the droplets where we expect He$_2^+$ and larger He cationic clusters to form. Note that the penetration depth of the XUV photons $1/(\sigma_\mathrm{abs}n_\mathrm{He})\gtrsim 1800$~\AA~is much larger than the mean He droplet size so that we may expect nearly uniform illumination of all He atoms in the droplets. Here, $\sigma_\mathrm{abs}=0.026$-$0.012$~\AA$^2$ is the photoionization cross section of He in the photon energy range $44$-$64$~eV~\cite{Samson:1994}. Our speculative explanation is that the bound He$^*$-He$^+$ pair of atoms is expelled towards the He droplet surface where it decays into an unbound pair of atoms He+He$^+$. In the bound excited state, the excited electron is delocalized over the two He atoms and therefore the system represents a vibronically excited molecular ion. Vibrationally excited~\cite{Callicoatt:1998,Smolarek:2010} as well as electronically excited molecular ions~\cite{Brauer:2011} and even excited atomic ions~\cite{Zhang:2012} have been found to be efficiently ejected out of He nanodroplets by a nonthermal, impulsive process. The He$^*$-He$^+$ system either detaches from the droplet, or it stays very weakly bound at the surface until it radiatively decays into an unbound pair of atoms He+He$^+$ with an interatomic spacing around 4~\AA~given by the minima of the potential wells (green and blue curves in Fig.~\ref{fig:Potentials}). 
Our observation of the disappearance of the He$^*$-He$^+$ molecular features for large He droplets is in line with previous observations of reduced ion yields for larger droplets~\cite{Smolarek:2010}.



\begin{figure}
	\centering \includegraphics[width=0.45\textwidth]{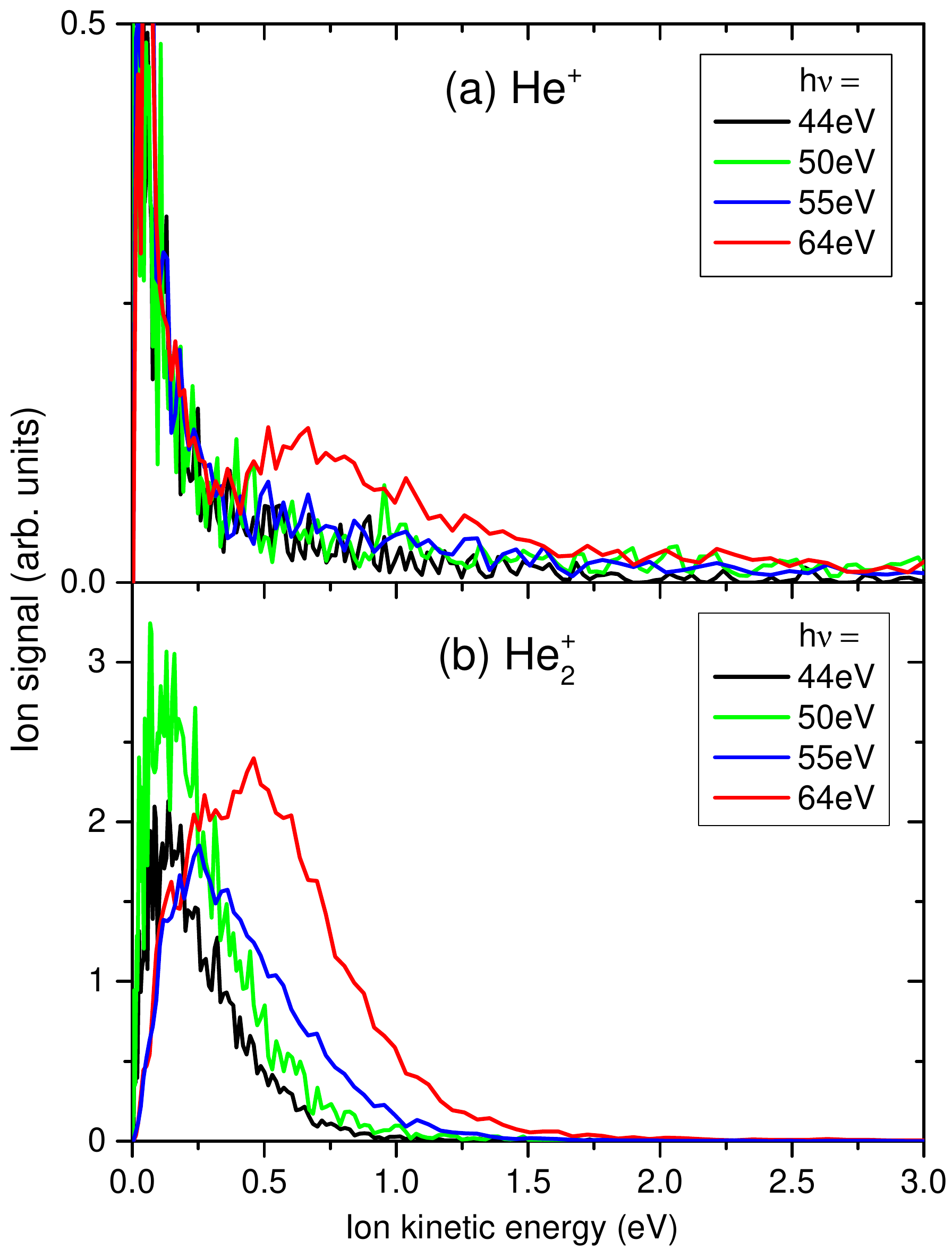}
	\protect\caption{\label{fig:ionKER} He$^+$ (a) and He$_2^+$ (b) ion kinetic energy distributions recorded at various photon energies.}
\end{figure}
Our interpretation in terms of inelastic scattering of the photoelectron with a He atom next to the atom from which it is emitted inside the droplet is supported by measurements of the He$^+$ and He$_2^+$ ion kinetic energy distributions, shown in Fig.~\ref{fig:ionKER} a) and b), respectively. In these experiments the He nozzle temperature was set to 16~K which corresponds to a mean number of He atoms per droplet of 5,500. At photon energies below the threshold for electron inelastic scattering, $h\nu=44$~eV, He$^+$ ions predominantly have low kinetic energies $<0.2$~eV. The kinetic energies of the He$_2^+$ cations are distributed around 0.15~eV with the highest energies reaching up to about 1~eV. The more extended low-energy distribution for He$_2^+$ ions as compared to He$^+$ is due to nonthermal ejection of vibrationally excited He$_2^+$~\cite{Buchenau:1990,Smolarek:2010}. At $h\nu=50$~eV, where various electron impact excitation channels open up, the He$^+$ and He$_2^+$ ion kinetic energy distributions are nearly unchanged. In contrast, at $h\nu=55$ and $64$~eV, where electron impact ionization is the dominant channel, the ion kinetic energy distributions qualitatively change; a second maximum appears as a shoulder in the ion spectrum of He$^+$ which is peaked around 0.7~eV and reaches up to about 2.7~eV. Similarly, the He$_2^+$ spectrum develops a shoulder peaked around 0.6~eV which extends up to about 1.7~eV. 

The higher kinetic-energy component of the ion energy distributions associated with electron impact ionization results from Coulomb explosion of pairs of ions created at relatively short distance. This is no surprise given the relatively large ionization cross section $\sigma_{i}=0.15$~\AA$^2$ at $h\nu=64$~eV~\cite{Ralchenko:2008}, which yields a probability of electron scattering with a neighboring He atom in a droplet at a distance of $d=3.6$~\AA~of $1-\exp(-\sigma_i n_\mathrm{He} d) = 1.2$\,\%. The Coulomb explosion of two adjacent atoms in a He nanodroplet spaced by 3.6~\AA~results in a kinetic energy release of about 4~eV, see the uppermost potential energy curve in Fig.~\ref{fig:Potentials}. This corresponds to a gain of kinetic energy of the He$^+$ ion of about 2~eV when assuming binary dissociation. 

The experimental higher kinetic-energy shoulder in the He$^+$ distribution at $h\nu=64$~eV contains 50\,\% of the total signal (25\,\% at $h\nu=55$~eV) and spans the wide energy range 0.35 - 2.7~eV. How can we rationalize this finding? The high-energy edge (2.7~eV) matches well the maximum energy that a He$^+$ ions can acquire when Coulomb explosion starts at the minimum distance between nearst neighbors (2.4~\AA)~\cite{Peterka:2007}. The low-energy onset (0.35~eV) corresponds to a distance between He$^+$ ions of about 20~\AA, which roughly coincides with the mean He nanodroplet radius (32~\AA). Thus, it seems that all electron-impact ionization events contribute to the shoulder structure in the He$^+$ energy distribution. Considering that scattering of the photoelectron with any atom in the entire He droplet may occur, we set $d=32$~\AA~and obtain an estimated scattering probability of $20$\,\%, which comes close to the experimental value. As in the case of electron-impact excitation of a neighboring atom discussed above, a correlated or even collective one-photon double ionization process akin to shake-off~\cite{Schneider:2002} may enhance the signal amplitude. 

The shape of the shoulder distribution is given by the distribution function of distances at which impact ionization occurs. In addition, collisions of the accelerated ions with neighboring He atoms on their way out of the droplet likely cause a shift towards lower energies. Such elastic ion-atom collisions have recently been observed for pairs of ions created by interatomic Coulombic decay (ICD) inside He droplets at an even shorter distance~\cite{Shcherbinin:2017}. When an ion collides elastically with a neighboring He atom it can lose all of its kinetic energy (head-on collision), in which case Coulomb explosion restarts at a larger distance and the final kinetic energy is reduced. An accurate modeling of the measured ion kinetic energy distributions would require a three-dimensional scattering simulation, which goes beyond the scope of this paper, though.

\subsection{Electron impact cross sections}
\label{sec:xsec}
\begin{figure}
	\centering \includegraphics[width=0.45\textwidth]{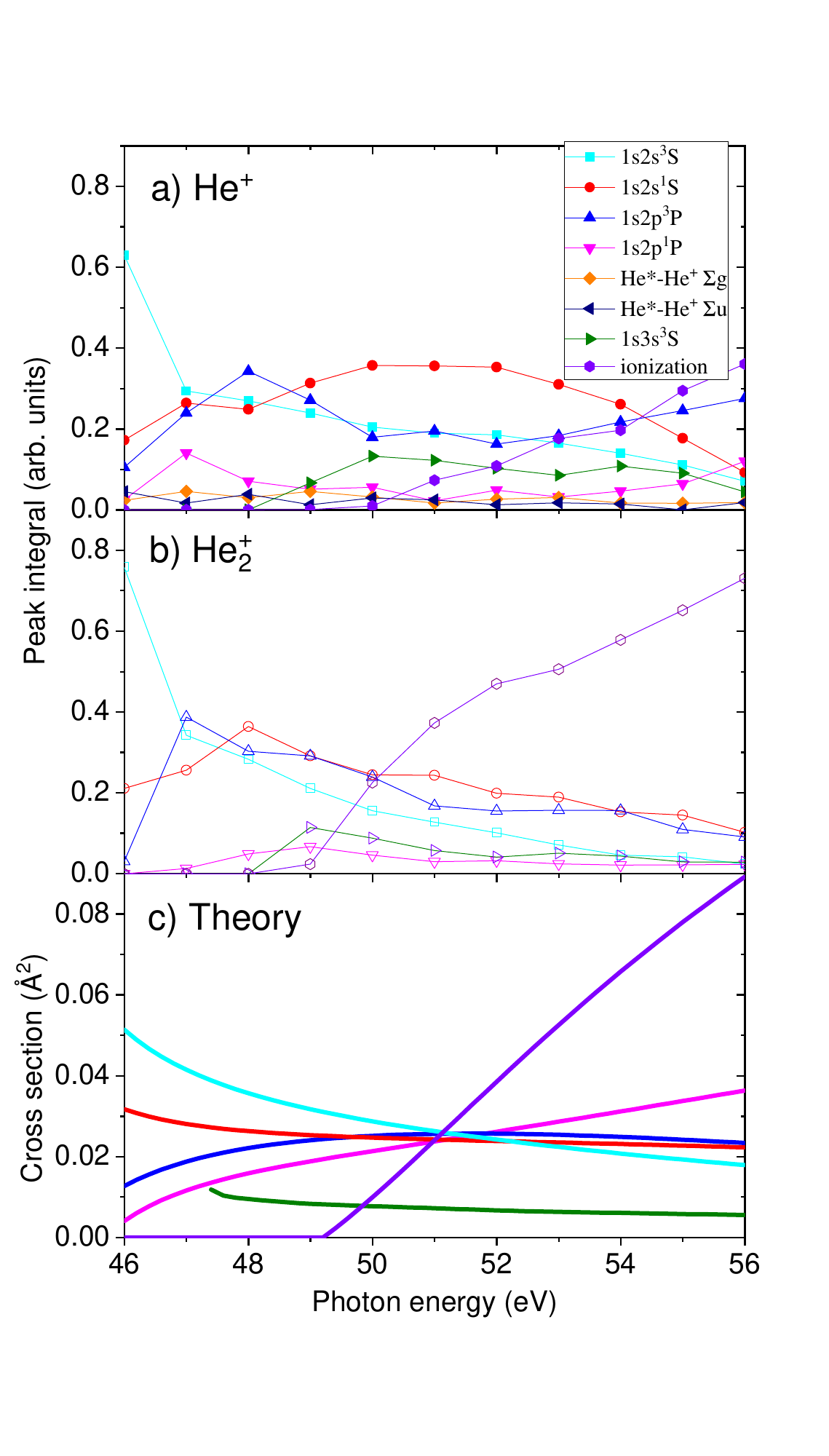}
	\protect\caption{\label{fig:Integrals} Integrals of the fitted peaks in the electron spectra recorded in coincidence with He$^+$ and He$_2^+$ [a) and b), respectively]. Panel c) shows the cross sections for electron impact excitation and ionization of various states of He extracted from Ref.~\cite{Ralchenko:2008}.}
\end{figure}
Finally, we inspect the relative intensities of the various inelastic electron scattering channels as a function of photon energy. Fig.~\ref{fig:Integrals} displays the integrals over individual fitted peaks in the electron spectra recorded in coincidence with He$^+$ (a) and with He$_2^+$ (b). For comparison, the theoretical results by Ralchenko~\cite{Ralchenko:2008} is shown in panel c). 
In the range $46\leq h\nu \leq 51$~eV the order of the experimental peak integrals roughly matches that of the theoretical cross section. Thus, 1s2s$^3$S excitation is most probable, whereas 1s2p$^1$P is least. At $h\nu > 51$~eV, the peak integrals are of similar order of magnitude, which agrees with the theoretical cross sections. 

Likewise, the predicted pronounced rise of the cross section for impact ionization for $h\nu >49.2$~eV is well reproduced by the electron data in coincidence with He$_2^+$. In the He$^+$ coincidence spectra, the impact ionization signals are slightly underrepresented with respect to the impact excitation channels.
Given that the ratio of detected He$^+$ ions and coincident electrons as compared to He$_2^+$ ions is only 0.23, the overall agreement of the measured peak integrals with the theoretical cross sections is satisfactory. This indicates that electron impact excitation and ionization proceed essentially with the same probabilities in He droplet as in free He atoms. 

\begin{figure}
	\centering \includegraphics[width=0.35\textwidth]{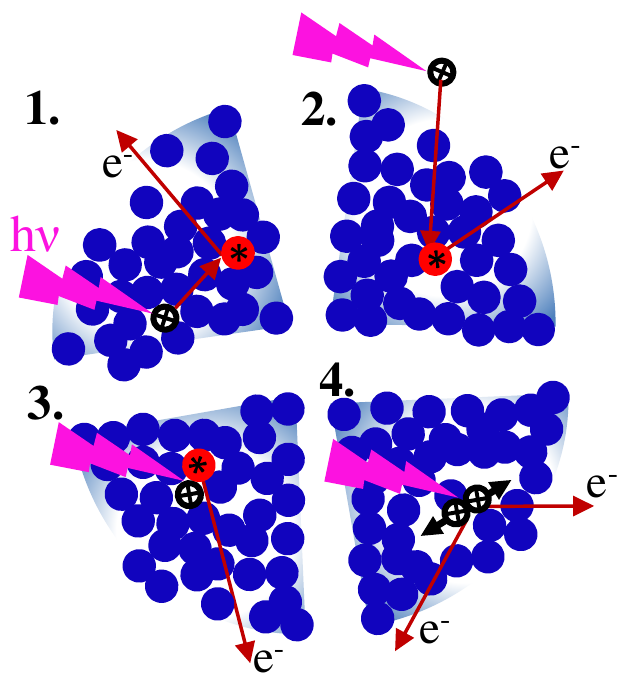}
	\protect\caption{\label{fig:Schematics} Schematic representation of four possible photoionization and electron-He inelastic scattering processes occurring upon photoionization of He nanodroplets at $h\nu >44$~eV. Excited He$^*$ atoms are marked by stars, He$^+$ ions are shown as circles containing crosses. See text for details.}
\end{figure}
\section{Conclusion}
When He nanodroplets are irradiated by XUV light at $h\nu\gtrsim 44$~eV, emitted photoelectrons undergo inelastic collisions with He atoms in the nanodroplets. The fraction of scattered electrons can exceed that of directly emitted electrons when the droplets grow large with radii $\gtrsim 20$~nm. The amplitudes of the individual inelastic channels (excited states of He, ionization) roughly agree with theoretical predictions for the He atom. 

Previously, we had concluded that photoionization events occurring inside the droplets mostly generate He$_2^+$ molecules and larger clusters in a process labeled by (1) in Fig.~\ref{fig:Schematics}. In contrast, He$^+$ ions mostly originate from photoionization of free He atoms that accompany the He droplet beam, see process (2). However, when photoionization takes place inside the droplet and the photoelectron scatters off a He atom in the close vicinity of the photoion, He$^+$ ions and correlated electrons originating from inside the He droplets can be detected as well. In the case that the next neighbor is excited into low-lying levels by collision with the photoelectron, a transient bound He$^*$-He$^+$ molecular state is populated [process (3)]. This is seen as two small peaks in the electron spectrum corresponding to excited levels lying below the lowest excited state of the He atom. In case that the next neighbor is impact ionized, the two He$^+$ ions undergo Coulomb explosion and both He$^+$ and He$_2^+$ ions are detected with higher kinetic energy of up to 3~eV [process (4)]. Thus, the two processes (3) and (4) are the pertinent new features related to He nanodroplets. The possibiliy that correlated or even collective effects contribute to the amplitudes of these processes is an intriguing thought that hopefully incites theoreticians to investigate this system.


These results add to the fundamental understanding of the interaction of relatively low-energy electrons with condensed phase systems. Electron scattering in biological matter plays a crucial role in radiation biology and DNA damage~\cite{Boudaiffa:2000}. Besides, our findings show that electron scattering may impose severe limitations for the use of He nanodroplets as a substrate for photoelectron spectroscopy of embedded molecules and complexes when using XUV or x-ray radiation. In a forthcoming study, we will discuss the importance of elastic scattering of electrons created inside large He nanodroplets.

\begin{acknowledgments}
M.M. acknowledges financial support by Deutsche Forschungsgemeinschaft (project MU 2347/10-1) and by Aarhus Universitets Forskningsfond (AUFF). A.C.L gratefully acknowledges the support by Carl-Zeiss-Stiftung. S.R.K. thanks DST, Govt. of India, for support."
\end{acknowledgments}


%

\end{document}